# The Decline and Fall of the Youngest Planetary Nebula

Bruce Balick[1*], Martín A. Guerrero[2], Gerardo Ramos-Larios[3]

[1] Department of Astronomy, University of Washington, Seattle, WA 98195-1580, USA
[2] Instituto de Astrofísica de Andalucía (IAA-CSIC), Glorieta de la Astronomía S/N, 18008 Granada, Spain
[3] Instituto de Astronomía y Meteorología, Universidad de Guadalajara, 44130 Guadalajara, Mexico
*Corresponding author: balick@uw.edu

**Abstract**

The Stingray Nebula, aka Hen3-1357, appeared for the first time in 1990 when bright nebular lines and radio emission that had not been observed before were unexpectedly discovered (Parthasarathy et al. 1993). In the ensuing years the nebula faded precipitously. We report changes in shape and large decreases in its nebular emission-line fluxes based on well-calibrated images obtained by the Hubble Space Telescope in 1996 and 2016. Hen3-1357 is now a "recombination nebula".

## 1. Introduction

Planetary nebulae ("PNe") consist of stellar gas and dust ejected in sustained and often structured winds from the surfaces of post Asymptotic Branch Giant ("AGB") stars. The ionization of the nebula typically unfolds over a millennium as the central star traverses a post-AGB track in the H-R Diagram (Schönberner et al. 2014, Miller Bertolami 2016).

The "Stingray Nebula" (aka Hen3-1357) had a sudden and very unusual and birth: although its post AGB star central star, SAO 224567 has been well-observed, the first evidence of the present nebula went undetected until it emerged from the 1980s as a typical compact yet fully ionized PN (Parthasarathy et al. 1993; hereafter "P+93")[1]. Presumably the nebular ionization began in the early 1980s when the B magnitude of SAO 244567, suddenly dropped by ≈ 0.5 mags—and continued to fade by nearly 1 mag per decade (Figure 1 in Schaefer & Edwards 2015; hereafter "SE15"), its surface temperature suddenly rose, and its radius dropped (Reindl et al. 2014; hereafter "R+14"). The line ratios in the first nebular spectrum obtained by P+93 were characteristic of those of many young PNe (P+93) with a hot central star ([O III]/H$\beta \gtrsim$ 10), as were the infrared colors of its central star (P. Garcia-Lario, 1992, PhD thesis).

The compact nebula (largest angular radius $\theta$ = 0.″8) was first imaged by Bobrowsky in 1992 using the then-new Hubble Space Telescope ("HST") (Bobrowsky, 1994; hereafter "B94", and soon confirmed in sundry narrowband images in 1996 (Bobrowsky et al. 1998, hereafter "B+98"). The images showed that the internal nebular density structure is highly articulated with large variations in surface brightness on scales of HST's spatial resolution, 0″.05. B94 identified Hen3-1357 as "the youngest PN" by B94, a title wrested from LMC SMO 64 (Dopita & Meatheringham 1991).

Judging from its present expansion speed, ~ 8.4 km s$^{-1}$ (Arkhipova et al. 2013; hereafter "A+13"), size $\theta$, and distance D, the present nebula was initially ejected ~1000y prior to the onset of ionization (R+14). Here estimates of D/kpc range from 0.83 (Fresneau et al., 2007), 1.6 (R+14), 1.8 (A+13) and 1.63–4.92 (Otsuka et al. 2017; hereafter "O+17"). So the time scale of gas ejection in Hen3-1357 is fairly standard. It is the ionization time scale, about a decade, that renders its apparition so remarkable.

---

[1] Soon confirmed Parthasarathy et al. (1995) and Feibelman, (1995), and Suarez et al. (2006)

We will show that the early brightening of the nebula almost immediately reversed course. We exploit high-resolution images of Hen3-1357 taken by HST between 1996 and 2016 to trace the trace its rapid changes in total emission-line fluxes and surface brightnesses during those 20 y (Figure 1). The total fluxes of the high-ionization emission lines H$\alpha$, H$\beta$, and [O III] were already in a state rapid decline in 1996, whereas the total fluxes low-ionization emission lines were dropping much more slowly or remaining constant. The nebular surface brightness measured from the calibrated HST images obtained during the 20-y interval reveals highly local decreases in the pixel count rates of all lines, albeit with a few outer zones *increasing* by factors up to three. We present the HST images used in our study, discuss the information extracted from them, and comment on the results in sections 2, 3, and 4, respectively.

**2. History**

Hen3-1357 has turned out to be the most rapidly evolving of all photoionized PNe. HST images (Figure 1) from 1996 and 2016 provide a glimpse into how its surface brightness, detailed shape, and the nebular ionization state fluctuated dramatically. Initial reports (Bobrowsky et al. 1998, A+13) suggested that the nebular emission line fluxes had changed erratically; i.e., the integrated low-ionization line ratios of [O I]/H$\beta$, [O II]/H$\beta$, and [N II]/H$\beta$ of Hen3-1357 had doubled between 1992 and 2011 whereas the [O III]/H$\beta$ ratio had decreased by a similar factor. Secular changes in its radio Bremsstrahlung continuum strength (Harvey-Smith et al. 2018) have been comparable in magnitude to the flux changes of Balmer lines of H$^+$, as expected. This confirms that foreground extinction is not the cause of the fading of Hen3-1357, in good agreement with the small interstellar reddening correction, $c$(H$\beta$) = 0.083, found by O+17. (See also R+14 for a summary of diverse stellar extinction results.)

SAO 244567 has also undergone rapid changes since about 1980. Several in-depth historical accounts of the evolution of the central star SAO 244567 exist (e.g., R+14, Schaefer & Edwards 2015; hereafter "BE15", and Reindl et al. 2017; hereafter "R+17"). In brief, the B magnitude of SAO 244567 faded steadily at a rate of 0.20 mag/y since nebular ionization began. The stellar surface temperature, $T_{\rm eff}$, rose from 21 kK by 1980 to 38 kK in 1988, to a peak value of 60 kK in 2002 and then cooled to 50 kK in 2015. Its surface gravity, log $g$, increased from 4.8 to 6.0 between 1988 and 2002 and then decreased to 5.4 in 2015 (Figure 4 in R+17) as the star contracted and then expanded. Unfortunately, there is no detailed record of the rapid changes of the fluxes and energy spectra of the UV photons that have been ionizing Hen3-1357 since the onset of ionization in the early 1980s.

The light travel time across Hen3-1357 is about a month. So, changes in the stellar UV photon fluxes above the ionization potentials of the emission-line species will be reflected almost immediately in the local ionization rates per ion, $I_{\rm ion}$, of each ion within the nebula. One the other hand, the recombination rates per ion, $R_{\rm ion}$, (and hence the local ionic abundances and emissivities) are considerably slower: $R_{\rm ion}$ = $10^5$ y/($Z^2 n_{\rm e}$), where Z is the ionic charge, $n_{\rm e}$ is the electron density, and where 6000 < $n_{\rm e}$ < 23,000 cm$^{-3}$ (O+17). There is little doubt that every part of this highly structured nebula has been out of ionization equilibrium, $I_{\rm ion} = R_{\rm ion}$, since nebular emission lines were first detected.

Finally, let us note that the nebular interior of Hen3-1357 almost certainly retains the shape and structure that it had when it was first ionized 30–40 y ago. Of course, ionization quicky heated its embedded filaments to ~$10^4$ K (O+17) causing them to thicken at their sound speed (~10 km s$^{-1}$). However, the thickening of the filaments will not yet be resolved in HST images; that is, the filaments would have thickened by ≲0″.01 in 20 years. More globally, such a thermally over-pressured nebula can expand and form a dense expanding rim with an ionization front (hereafter "IF") of D-type when $I_{\rm ion} \gg R_{\rm ion}$, likely with a dense rim and radiative shock along its perimeter if the early expansion was highly supersonic.[2]

---

[2] See Chapter 7 of Dyson and Williams (1981) for a discussion of the applicable physics.

## 3. Archival Results

Table 1. Detected Fluxes of Narrowband Images of Hen3-1357

| Integrated Line Fluxes ($10^{-12}$ erg cm$^{-2}$ s$^{-1}$) | H$\beta$[a] 486 nm | [O III][a] 501 nm | He I 587 nm | [O I] 630 nm | H$\alpha$ 656 nm | [N II] 658 nm | [S II] 672+3 nm |
|---|---|---|---|---|---|---|---|
| Camera Filter Name | F487N | F502N | F588N | F631N | F656N | F658N | F673N |
| 1996 (WFPC2) | 1.65 | 12.1 | 0.31 | 0.28 | 5.40 | 1.76 | 0.22 |
| 2016 (WFC3) | 0.67 | 2.45 | | | 1.99 | 1.03 | 0.21 |
| Flux ratio 2016/1996 | 0.41 | 0.20 | | | 0.37 | 0.59 | 0.96 |

a. Extinction-corrected (H$\beta$, [O III]) fluxes were (1.9 x $10^{-11}$, 2.2 x $10^{-9}$) erg cm$^{-2}$ s$^{-1}$ in 1990 and (5.8 x $10^{-11}$, 6.4 x $10^{-10}$) erg cm$^{-2}$ s$^{-1}$ in 1992 (B94).

Table 2. Flux Ratios of Image Total Fluxes to H$\beta$[a]

| Line Fluxes/H$\beta$ | (H$\beta$=1) | [O III]/H$\beta$ | He I/H$\beta$ | [O I]/H$\beta$ | H$\alpha$/H$\beta$ | [N II]/H$\beta$ | [S II]/H$\beta$ |
|---|---|---|---|---|---|---|---|
| *1990 (P+93)* | *(1)* | *~9[b]* | *0.16* | *0.04* | *3.0* | *0.8* | *0.05* |
| 1996 (WFPC2) | (1) | 7.3 | 0.19 | 0.17 | 3.3 | 1.1 | 0.13 |
| *2006 (O+17)* | *(1)* | *4.2* | *0.15* | *1.6* | *(3.0)[c]* | *1.2* | *0.18* |
| 2016 (WFC3) | (1) | 3.7 | | | 3.0 | 1.5 | 0.31 |

a. Entries in italics come from spectra as published in the references shown in the first column
b. Derived from 3 × [O III]l4959/H$\beta$ = 3.05 in Table 1 of P+93.
c. Estimated (see discussion in O+17)

The HST images of Hen3-1357 used in this paper were obtained using two telescope cameras, WFPC2 and WFC3, each with its own various filters and detectors—and downloaded from the Hubble Legacy Archive[3] ("HLA"). We divided the count rates of each downloaded image by the total throughput efficiencies corresponding to its camera-filter combination. The veracity of their relative intensities was confirmed to ±10% using non-variable field stars after applying corrections for the bandwidths of the filters. The images were transformed to the same pixel size and then spatially registered using the central star, or if it was saturated in the image, its fainter and cooler visual companion about 1″ to the northeast

Figure 2 contains a montage of the image ratios of Hen3-1357 between 1996 and 2016 in [O III], [N II], and H$\alpha$, respectively. The emission lines and corresponding filters are listed across the top. Upper row (left to right): ratios of the calibrated images (2016/1996), where white pixels within the nebula repre–sent absolute count-rate ratios of 0.5, 2.0, and 0.6 (black is zero and ignore the central star.) Loci of exceptional values are indicated by their values and small arrows. The bottom row shows the images after their peak pixels were normalized to unity. The 1996 (2016) images are shown in orange (cyan). These images emphasize changes in the nebular shapes and internal surface brightnesses.

Figures 1 and 2 illustrate that all four of the low-ionization lobes and edge arcs that were highly prominent in 1996 faded dramatically by 2016. The central elliptical ring of Hen3-1357 also faded at a far greater rate in [O III] and H$\alpha$ than in it did in [N II]. Note also the apparent shrinkage of the minor axis of the central ellipse. [S II] images (HST filter F673N, not shown) emulate the same 20-y changes as [N II] except somewhat more slowly. There is no evidence of any systematic expansion of the nebula, consistent with the nebula's expansion speed of 8.4 km s$^{-1}$ (A+13).

The calibrated pixel count rates of each image were converted to fluxes, erg cm$^{-2}$ s$^{-1}$ pixel$^{-1}$ and summed over the nebular image in order to measure total nebular line fluxes. The results are compiled in Table 1. Table 2 shows the ratios of these fluxes relative to H$\beta$ (= 1) from the 1996 images (top row) and 2016 images (middle row). Other spectrophotometric estimates of the ratios are shown in italics. We emphasize that the ratios of total line fluxes of [N II]/H$\beta$ and [S II]/H$\beta$ increased over 20 y whereas [O III]/H$\beta$ decreased by 50%. The temporal trends are steady.

---

[3] Based on observations made with the NASA/ESA Hubble Space Telescope, and obtained from the Hubble Legacy Archive, which is a collaboration between the Space Telescope Science Institute (STScI/NASA), the Space Telescope European Coordinating Facility (ST-ECF/ESA) and the Canadian Astronomy Data Centre (CADC/NRC/CSA).

Finally, we used the 1996 images to make images of the ratios of the forbidden lines to permitted lines at about the same wavelength in order to investigate the ionization structure early in the life of the nebula. The image ratios are shown in Figure 3 in order of increasing ionization energies, [O I]/H$\alpha$, [S II]/H$\alpha$, [N II]/H$\alpha$, and [O III]/H$\beta$. All of these ratios rise abruptly at the nebular perimeter and tend to be smallest in internal features of high emission measure (i.e., density).

## 4. Discussion

Nebular ionization in 1996. We first discuss the early ionization state of Hen3-1357 before we explore its subsequent changes. Large-aperture optical spectra from 1990 (P+93), 2003 (Suarez et al. 2006) and 2006 (O+17) show emission line ratios that are characteristic of nearly all photoionized PNe (Frew & Parker, 2010). In addition, Figure 1 shows that the ionization state of the gas decreases from the inside out, as expected from UV photoionization models (Ferland et al. 2013). Thus there is no doubt that stellar UV ionization has dominated the average ionization state of the gas since 1990 or possibly before.

However, the line ratios that were seen at the nebular margins in 1996 (Figure 3) raise complex and interesting questions. The peak line ratios of [O I]/H$\alpha$, [S II]/H$\alpha$, [N II]/H$\alpha$, and [O III]/H$b$ are larger by factors of 110, 90, 3, and 2 compared to the corresponding whole-nebula flux ratios measured by P+93 in 1990 using a slit spectrum taken in unspecified seeing (Table 2). The two most plausible explanations for these large discrepancies are (1) an IF at the interface between the ionized nebula and external neutral gas and (2) an abrupt local rise in the local forbidden-line excitation temperatures, $T_{exc}$, along the edge (that is, optical forbidden lines are readily excited by thermal collisions at $T_{exc} \gtrsim 10^{3.7}$ K).

The highly filamentary density structure inside Hen3-1357 mitigates against the first alternative since the presence of a smooth and continuous IF along a highly articulated nebular boundary cannot be expected. The second alternative might apply if collisional shocks arise along the outer edge of Hen3-1357. As already noted, the early expansion of the highly over-pressured hot ionized gas (when $I_{H+} >$ $R_{H+}$) may initiate a supersonic blast wave that propagates into and shocks the cold ambient gas even after $I_{H+} < R_{H+}$. We also note that the values of the line ratios at the nebular edges generally agree with wind-shock models for H-H objects (e.g., Raga et al. 2008), especially the large [O I]/H$\alpha$ and [S II]/H$\alpha$ ratios. Further support comes from the symmetric shapes of each of the two pairs of opposing lobes, two of which ($L_{SE}$ and $L_{NW}$) resemble bow shocks formed by narrow jets. Accordingly, we can surmise that the low-ionization perimeter of Hen3-1357 is shock excited. However, given the absence of nebular ionization equilibrium throughout Hen3-1357 we best not jump to conclusions.

Nebular ionization since 1996. To understand the subsequent changes in nebular brightness, color, and shape of Hen3-1357, first recall that $R_{ion}$ scales as $\approx 10^5$ y/($Z^2 n_e$). For example, the O$^{++}$ recombination time scale, $\tau_{O++}$, is 1.3 y if $n_e \sim 2 \times 10^4$ cm$^{-3}$. Thus, the integrated flux of the [O III] line dropped the fastest—by a factor of 900 since 1990—owing to the relatively rapid recombination rate of O$^{++}$ ions in the densest, most emissive regions of the nebula. The H$\beta$ flux dropped by a much smaller factor of 29 during the same interval, (not 900/4 probably because the average densities in the H$^+$ volume are smaller than those in the O$^{++}$ ionization zone (O+17)). On the other hand, the fluxes of the N$^+$ and S$^+$ decreased much less between 1996 and 2016. This is likely because recombinations of N$^{++}$ and S$^{++}$ enhanced the ionization fractions of N$^+$ and S$^+$ faster than the recombinations depleted them. In any event, it is clear that Hen3-1357 is presently a "recombination nebula" in which the local and volumetric rates of $I_{ion}$ and $R_{ion}$ are out of balance.

The complex effects of recombinations on the nebular surface brightnesses of [O III], [N II], and H$\alpha$ within Hen3-1357 are shown in the top row of Figure 2. These images show the ratios of the 2016 to 1996 images in these lines, where the peak values of the respective ratios are seen in white and the zero level is black. Let us define the nebular "core" as the zone inside the lobes and edge arcs where the H$\alpha$ and [O III] lines dominated the 1996 images. The fading patterns inside the core are much more complex than outside, although it is clear they are closely related to the internal density structure of the nebula.

The differences of the local fading rates among the [N II], H$\alpha$, and [O III] lines are especially interesting. In 1996 the most prominent [N II] features of Hen3-1357 were the edge arcs and the lobes L$_{NE}$ and L$_{SW}$. However, these features nearly vanished by 2016, presumably as recombinations cooled the gas and depleted the local fraction N$^+$ ions. In contrast, the brightness of [N II] increased inside the elliptical ring as recombinations of N$^{++}$ enhanced the local fraction of N$^+$ in the densest structures (with only minor effects on the local value of $T_{exc}$).

H$\alpha$ faded more somewhat more uniformly across the nebula than did the forbidden lines. The behavior of H$\alpha$ is the simplest to understand since the forbidden line recombination rates are much more sensitive to the local value $T_{exc}$ as well as recombinations of doubly ionized species. The slowest fading of H$\alpha$ occurs along the inside edge of the elliptical ring—in regions immediately outside of the filaments at the boundary of the core—and just beyond the tips of L$_{SE}$ and L$_{NW}$. It seems that $n_e$ is relatively small within these regions just beyond the core's boundaries.

[O III] faded everywhere and most rapidly outside the core. The simplest explanation for this is that $T_{exc}$ in these zones dropped thanks in part to local cooling of the gas as the zone became neutral. However, some [O III] emission is found beyond the tips of lobes L$_{NW}$ and L$_{SE}$ where some shock heating from winds or nebular expansion may have persisted.

**5. Final Remarks**

The highest aspiration for a study such as this one could be to constrain conflicting models for the erratic evolution of SAO 244567 (see SE15). To explain its behavior the Reindl collaboration group (R+14, R+17) proposed that the star underwent a disruptive late thermal pulse ("LTP") perhaps as long as forty years ago and is now returning to the top of the Asymptotic Giant Branch in the H-R Diagram. However, we note SE15 presented alternate interpretations for the present state of the star with which R+17 take exception. Unfortunately, our data cannot address these important issues.

Firstly, we probe how the properties of the nebula might inform stellar models for central stars that evolve rapidly. Observations of the ionized gas of Hen3-1357 fail to place significant insights on the behavior of SAO 244567. On the one hand, the ionization state of Hen3-1357 probably responded very quickly to sudden changes in the ionizing photon flux and energy spectrum of SAO 244567 during the 1980s. Unfortunately, the nebula was not observed until much later by which time electron-ion recombinations of sundry ions within the complex density structure of the nebular gas had significantly modified its initial ionization state. That is, the response of the nebula, no matter how well observed, is dominated by recombination time scales that for each ion, $R_{ion}$, that lag the variations in $I_{ion}$ on very diverse time scales within its complex density and ionization structure. Of course, we can safely conclude that some combination of ionizing photon flux and energy has also declined since 1990 (if not even earlier). However, using the ionic emission lines to study the star is much like using smoke to study the progress of a forest fire. Within an H-R Diagram, "The position of [an] LTP or VLTP [star] cannot be inferred from nebular line analysis" (D. Schönberner, private communication).

Secondly, we briefly explore what can we learn about the star and the nebula from their analogues. Hen2-47 (aka MY 59) is a very young "starfish PN" that closely resembles to Hen3-1357 in form, ionization, and kinematics. Like Hen3-1357, Hen2-47 has a bright inner elliptical ring and multiple pairs of outer lobes. Its central star, Gaia DR2 5254793942926363392, is spectral type O(H)7-9 I and exhibits Balmer and HeII absorption lines (Weidmann et al. 2018). A spectrum of Hen2-47 from 2004 shows bright Balmer, [O II], and [N II] lines and faint [O III] lines (see "The Macquarie/AAO/Strasbourg Hα Planetary Nebula Catalogue: MASH", Parker et al. 2006). Otherwise, Hen2-47 has been far less observed than Hen3-1357 with little information on its evolving properties.

We obtained and divided F656N images of Hen2-47 obtained in 1999 and 2007 from the Hubble Legacy Archive. The east-west pair of lobes faded by ~4% over eight years while the brightness of the rest of the nebula and its central star remained constant. There is suggestive but inconclusive evidence that the central elliptical ring and the overall nebula increased in size in ten years. Thus, despite its nebula changes in surface brightness, Hen2-47 is not particularly useful for studying the evolution of its central star or informing our study of Hen3-1357.

There are other ionized nebulae near highly evolved stars that have evolved more quickly than expected from standard post-AGB stellar evolution theory. Most notably, HuBi 1is a *bone fide* PN that, like Hen3-1357, is actively recombining after its central star dimmed by ~10 mag in the last 50y (Guerrero et al. 2018, Rechy-García, et al. 2020). The time-evolution of fluxes of HuBi 1's emission lines behave roughly similarly to Hen3-1357, albeit on time scales ~5 longer than those of Hen3-1357. Other less studied examples include HDW 11, K 2−2, GD 561, PHL 932, and DeHt5, though most of these are not true PNe (Gunther Cibis, private communication).

There are very few highly evolved stars that have shown rapid variability on about the same time scales as SAO 244567. Let us first note that R+14 assert that the evolution of SAO 244567 is a result of a late thermal pulse ("LTP"), not a very late thermal pulse ("VLTP")[4]. Like Hen3-1357, the emission lines and ionization state of Sakurai's object (V4334 Sgr) declined on a time scale of 2 y between 2002 and 2006 following a VLTP in the central star. However, V4334 Sgr is a H-depleted VLTP star (R+14, SE15). Van Hoof et al. (2007) attribute this decline to recombinations and cooling in a post-shock zone (not a photoionized nebula). Therefore, both the nebula and its central star make poor analogies of Hen3-1357 and SAO 244567, respectively.

Similarly, V605 Aql (aka PN 58, Nova Aql 1919) is a star with a record of very rapid decline following a VLTP or nova outburst. Wesson et al. (2008) showed that a recently ejected and extremely hot nebular knot is rich in O and Ne and highly deficient in H. Another possible analogue is FG Sge, star within a nebula like Hen3-1357 that underwent an LTP event centuries ago. However, its nebula is ten times larger (and may be far older) than Hen3-1357. Additionally, its nebular evolution time scale is about 100 times slower than that of Hen3-1357 (Lawlor & McDonald 2003). Interestingly, FG Sge returned to the AGB and has become hydrogen deficient. Thus, the major discrepancies among Hen3-1357 and its most probable analogues provide no insight into the odd behaviors of SAO 244567 or its nebula Hen3-1357.

**Acknowledgements**

Acknowledgements: MAG acknowledges financial support by grant PGC2018-102184-B-I00, co-funded with FEDER funds. GR-L acknowledges financial support from CONACyT grant 263373. We thank D. Schönberner for very useful insights. B.B. thanks Gunther Cibis for informing him that although they are surrounded by ionized gas, K 2−2, GD 561, PHL 932, and DeHt5 are probably not true PNe.    This research has made use of the HASH PN database at hashpn.space.

---

[4] See R+14 for a brief discussion and comparison of these phenomena and their observable consequences.

**Facility** HST (WFPC2, WFC3)

**References**

Arkhipova, V.P., Ikonnikova, N.P., et al. 2013AstL...39..201A
Bobrowsky, M. 1994ApJ...426L..47B
Bobrowsky, M., Sahu, K.C., et al. 1998Natur.392..469B
Dopita, M.A. and Meatheringham, S.J. 1991ApJ...374L..21D
Feibelman, W.A. 1995ApJ...443..245F
Ferland, G.J., Porter, R.L., et al. 2013ApJ...767..123F
Fresneau, A., Vaughn, A.E., and Argyle, R.R.  2007A&A...469.1221F
Frew, D.J. & Parker, Q.A. 2010PASA...27..129F
Guerrero, M., Fang, X., et al. 2018NatAs...2..784G
Harvey-Smith, L., Hardwick, J. A., et al. 2018MNRAS.479.1842H
Lawlor, T. M., & MacDonald, J. 2003, ApJ, 583, 913
Miller Bertolami, M.M., 2016A&A...588A..25M
Otsuka, M., Parthasarathy, M., et al. 2017ApJ...838...71O
Parker, Q.A.,  Acker, A., et al. 2006MNRAS.373...79P
Parthasarathy, M., Garcia-Lario, P., et al. 1993A&A...267L..19P
Parthasarathy, M., Garcia-Lario, P., et al. 1995A&A...300L..25P
Raga, A.C., Riera, A., et al. 2008A&A...489.1141R
Rechy-García, R.S., Guerrero, M.A., et al., 2020, to be published in ApJ Letters.
Reindl, N., Rauch, T., et al. 2014A&A...565A..40R
Reindl, N., Rauch, T., et al. 2017MNRAS.464L..51R
Sahai, R. 2000ApJ...537L..43S
Schaefer, B.E. & Edwards, Z. 2015ApJ...812..133S
Schönberner, D., Jacob R., et al. 2014AN....335..378S
Schönberner, D., Jacob R., & Stefen, M. 2005A&A...441..573S
Suárez, O., García-Lario, P., et al. 2006A&A...458..173S
Van Hoof, P.A.M., Hajduk, M., et al. 2007A&A...471L...9V
Weidmann W.A., Ganen,, R. et al. 2018A&A...614A.135W
Wesson, R., Barlow, M. J., et al. 2008MNRAS.383.1639W

**Figures**

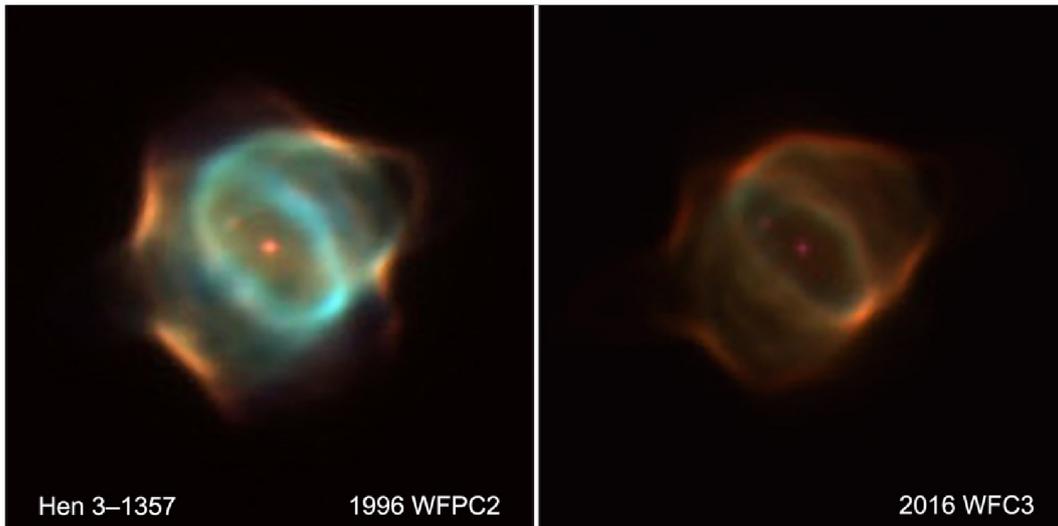

Figure 1. Color-composite F658N (red), F656N (green) and F502N (blue) *Hubble Space Telescope* images of Hen3-1357 taken in 1996 and 2016. The intensity levels of the two panels are the same, thus revealing the changes in surface brightness in the emission lines of $N^+$ (red), $H^+$ (green) and $O^{++}$ (blue).

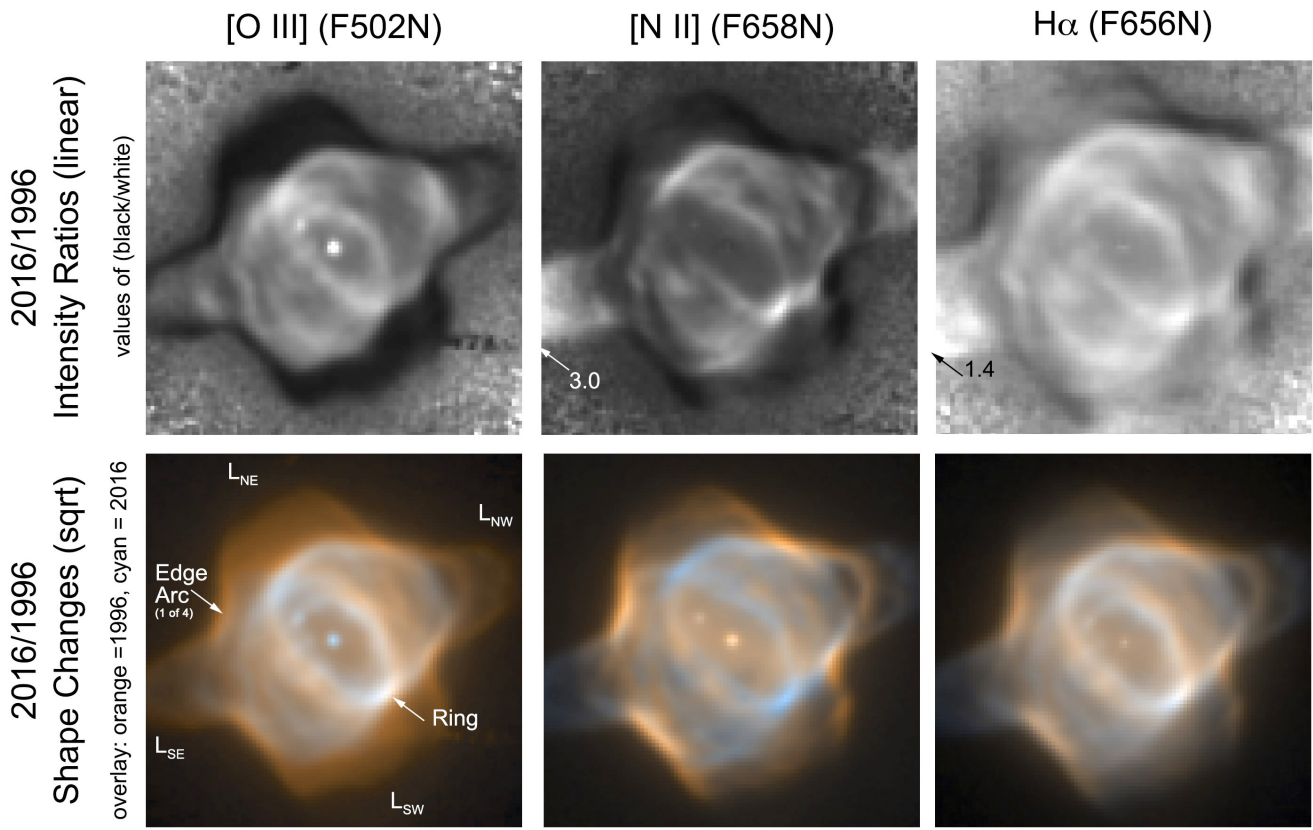

Figure 2. A montage of the image ratios of Hen3-1357 between 1996 and 2016 in [O III], [N II], and H$\alpha$, respectively. The emission lines are listed across the top. The upper row shows the ratios of the absolute count rates per pixel (nebular surface brightness) over 20 y, where white represents values of 0.5, 2.0, and 0.6 from left to right and black is zero. The bottom row shows the 1996 and 2016 images are shown in orange and cyan, respectively, after each image was renormalized to a peak count rate of unity. These overlays reveal how the ionization structure of the images changed in each emission line.

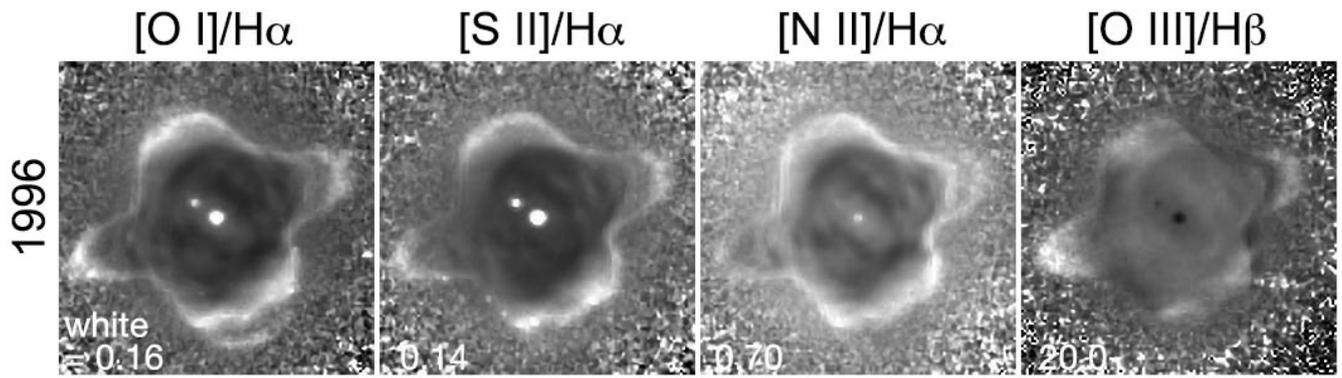

Figure 3. The ratios of various forbidden and permitted line images of Hen3-1357 observed by HST in 1996. The values corresponding to white in each panel are shown in their lower left corners.